\documentstyle[prl,aps,preprint,amssymb,psfig]{revtex}

\begin{document}
%
%
\draft
\title{Heat Capacity of MgB$_2$: Evidence for Moderately Strong
Coupling Behavior}

%
%
\author{R.K.\ Kremer, B.J.\ Gibson, and K. Ahn}
%
%
\address{Max-Planck-Institut f\"ur Festk\"orperforschung, Heisenbergstr. 1,
  D-70569 Stuttgart, Germany}
\date{\today}
\maketitle
%
%
\begin{abstract}
We characterize the superconducting state of phase pure
polycrystalline samples of the new layered superconductor MgB$_2$
by specific heat measurements in magnetic fields up to 9 T. The
characteristic jump at the superconducting transition is observed
and compared with  the predictions of weak coupling BCS-theory and
the $\alpha$-model. Our analysis shows excellent agreement with
the predictions  for 2$\Delta/k_BT_{\rm C}$\,=\,4.2(2) with a
Sommerfeld term $\gamma_{\rm exp}$ of 3.1(1) mJ/mol\,K$^2$
indicating that MgB$_2$ is a superconductor in the moderately
strong electron-phonon coupling regime.
\end{abstract} 

%
%

Recently, Akimitsu and co-workers reported superconductivity  at
$T_{\rm C}\approx$ 39\,K in the layered diboride MgB$_2$
\cite{Akimitsu}. This discovery on the one hand  again focuses
attention on the borides as possible candidates for high-$T_{\rm
C}$ superconductivity. On the other hand, $T_{\rm C}$ of MgB$_2$
exceeds the generally agreed theoretical limit for phonon mediated
superconductivity and raises the question as to another possible
coupling mechanism \cite{Ginzburg}. However, first measurements of
the thermodynamic properties and in particular of a sizeable
partial $^{11}$B isotope effect by Bud'ko et al. strongly hint at
the importance of phonons for superconductivity in MgB$_2$
\cite{Budko}. Kortus and co-workers, based on electronic band
structure calculations conclude that the metallic character is due
to covalent B-B bonding and that $T_{\rm C}$ particularly benefits
from {\it strong} electron-phonon coupling in concert with high
frequency vibrations associated with the light mass of the boron
atoms \cite{Kortus}. Evidence for strong-coupling $s$-wave
superconductivity was indeed found from $^{11}$B nuclear spin
lattice relaxation measurements from which a rather large gap
ratio of 2$\Delta$/$k_BT_C \approx$ 5 was derived \cite{Kotegawa}.
Strong electron-lattice coupling also seems to be able to explain
Al doping experiments carried out by Slusky et al. that show a
decrease of $T_{\rm C}$ with increasing Al content and finally a
loss of bulk superconductivity at $\approx$\,10\,\% Al doping
\cite {Slusky}. These experiments, in particular,  reveal  MgB$_2$
to be close to a structural instability involving a boron
interlayer alternation rather than bond alternation in the B
layers.
 All presently available tunneling spectroscopy experiments
consistently fit very well to an $s$-wave BCS quasi-particle
density of states and consistently exclude $d$-wave symmetry of
the order parameter. However, the gap values resulting from these
studies span a wide range and currently leave a somewhat
inconclusive situation. The very first experiments by
Rubio-Bollinger et al. gave a surprisingly small value of 2 meV
(weak-coupling BCS value is 5.9 meV) \cite{Rubio}. More recent
work by Schmidt et al. \cite{Schmidt} ($\Delta$=4.3 meV) and
Sharoni et al. \cite{Sharoni} ($\Delta$=5 - 7 meV) put $\Delta$
closer to the weak-coupling value or even into an {\it
intermediate coupling} regime. This wide range of gap values seems
to result from defects or minor non-superconducting impurity
phases, or chemical reactions at the surface of the
polycrystalline specimen.

In this Letter, we report a heat capacity study in zero field and
a magnetic field of 9 T of phase pure samples of MgB$_2$. In the
zero-field measurements we observe the characteristic jump in
$c_P$ at the superconducting transition temperature $T_{\rm
C}$=38.5\,K associated with the formation of the superconducting
condensate. Our $\Delta c_P (T_{\rm C})$ is in good agreement with
the finding of Bud'ko et al. \cite{Budko}. However, a more
detailed analysis of the anomaly was not carried out in their
work. The detailed temperature dependence  of the heat capacity
anomaly fits very well to  model calculations assuming a BCS-like
temperature dependence of the gap and a gap ratio
2$\Delta(0)$/$k_B\,T_{\rm C}\approx4.2(2)$. This value is
significantly increased over the BCS ratio of
2$\Delta(0)$/$k_B\,T_{\rm C}$=3.52. From the low temperature data
taken at 9 T we extract a Sommerfeld coefficient $\gamma_{\rm
exp}$\,=\,3.1(1) mJ/mol\,K$^2$. Comparing this value with the
results of band structure calculations we derive an
electron-phonon coupling parameter $\lambda_{\rm
el-ph}\approx$\,0.8 in good agreement with theoretical estimates
from linear-response electronic structure calculations
\cite{Kong}. Our findings for $\gamma_{\rm exp}$ and $\Delta
c_P(T_{\rm C})$ imply a ratio $\Delta c_P(T_{\rm C})/\gamma_{\rm
exp} T_{\rm C}\approx$0.7 which is significantly lower than e.g.
for the weak coupling BCS value of 1.43.

Polycrystalline samples of MgB$_2$ were prepared from
stoichiometric mixtures  according to the procedure described in
ref. \cite{Akimitsu}. Phase purity was confirmed with a STOE
powder diffractometer. The superconducting transition temperatures
$T_{\rm C}$ were determined with a Quantum Design MPMS7
magnetometer and conventional 4-point resistivity measurements and
amounted to $\approx$38.5 K. For the heat capacity measurements
pellets of $\approx$3 mm diameter and $\approx$15 mg were pressed,
sealed in Ta tubes under Ar atmosphere, and sintered at 950 $^{\rm
o}$C (sample 1) and 850 $^{\rm o}$C (sample 2) for 10 h. The heat
capacity was measured with a Quantum Design PPMS relaxation
calorimeter in the temperature range 1.8 to 100\,K in fields up to 9
T. Particular care was taken to determine the addenda and thermal
relaxation parameters of the platform and the suspension wires at
each magnetic field individually \cite{addenda}.

Fig.\,1 displays the zero-field heat capacity of MgB$_2$ in
comparison with the data taken within an external field of 9 T. An
anomaly centered at $\approx$38 K is visible in the zero-field
data which is suppressed by a magnetic field of 9 T. The maximum
of the anomaly amounts to $\Delta c_P(T_{\rm C})/T_{\rm C}\,\approx$ 2 mJ/mol
K$^2$. The difference of the zero-field and the heat capacity at 9
T is shown in Fig. 2 together with a theoretical curve (solid line
in Fig. 2). The theoretical calculations were carried out within
the framework of the `$\alpha$-model' proposed by Padamsee et al.
\cite{Padamsee}. This model assumes a BCS-type gap but allows a
variable ratio $\alpha\,\equiv\,\Delta(0)/k_{\rm B}\,T_{\rm C}$;
in the case of weak-coupling BCS theory $\alpha$\,=\,1.76.
 This simplistic approach was initially used to
model the specific heat jumps of elemental superconductors like
In, Sn and Zn, Pb - In alloys, and more recently also of layered
organic superconductors as well as of the layered rare earth
carbide halides \cite{Padamsee,Wosnitza,Henn}.

To calculate the heat capacity we used a polynomial representation
of the temperature dependence of the BCS gap as tabulated by
M\"uhlschlegel for $T\,\leq\,T_{\rm C}$ \cite{BCS}. Following
Padamsee et al. $\Delta c_P (T)$ was obtained according to

\begin{equation}
\Delta c_P/(\gamma_{\rm fit}\,T_{\rm C}) = C_{\rm
es}(t)/(\gamma_{\rm fit}\,T_{\rm C}) - t = d/dt\,(S_{\rm
es}/\gamma_{\rm fit}\,T_{\rm C}) - t
\end{equation}

with $t\,\equiv\,T/T_{\rm C}$ and the entropy $S_{\rm es}$ as
defined in eq.\,(2) in ref.\cite{Padamsee}. Subtraction of
high-field data from zero-field heat capacities to correct for the
lattice heat capacity contributions also implies a subtraction of
the normal state linear heat capacity term which is accounted for
by the -$t$ term in eq.\,(1). With $\Delta (T)\,\equiv\,$0, for
$T\,>\,T_{\rm C}$ eq.\,(1) leads to

\begin{equation}
\Delta c_P/\,T_{\rm C}\equiv 0.
\end{equation}

Fitting eq.\,(1) to
the experimental results therefore provides $\gamma_{\rm fit}$
merely from the magnitude of the anomaly. The detailed temperature
dependent shape of the anomaly is determined by $\alpha$. To fit
the experimental data we varied as adjustable parameters $\alpha$,
$\gamma_{\rm fit}$, $T_{\rm C}$, and  a parameter that simulated a
Gaussian broadening of $T_{\rm C}$ \cite{Henn}. In addition, we
allowed for a small shift of the baseline which never exceeded 2~
mJ/mol\,K.

The fits of the experimental results of the two independent samples
converge rapidly to $\alpha$ = 2.1(1) with $T_{\rm C}$=38.67(5)\,K
and $\delta T_{\rm C}$=0.55(5)\,K (sample 1) and with $T_{\rm
C}$=37.84(5)\,K and $\delta T_{\rm C}$=0.56(5)\,K (sample 2). $\delta
T_{\rm C}$ corresponds to $\sigma$ in a Gaussian distribution of
$T_{\rm C}$ and indicates a smearing of $T_{\rm C}$ of less than
1.5\,\% emphasizing good  homogeneity of both samples. The fits
consistently converged to $\gamma_{\rm fit}$=1.1(1) mJ/mol K$^2$
for both samples.

A reliable independent determination of the Sommerfeld coefficient
e.g. from low temperature heat capacity data turned out to be
difficult since a field of 9 T is not sufficient to suppress
superconductivity completely \cite{Takano,Budko2}. $H_{\rm c2}$
measurements indicate a critical temperature of $\approx$15K at 9
T. In a plot $c_P/T$ versus $T^2$ we observe a good linear
correlation already above 10K and marked deviations only below
$\approx$5\,K (see Fig. 3).

A fit of the 9 T data using $c_P/T$=$\gamma_{\rm exp}$ + $\beta
\cdot T^2$ yields $\gamma_{\rm exp}$=3.1(1) mJ/mol K$^2$ and
$\beta$=1.24(6)$\times$10$^{-5}$ J/mol K$^4$ corresponding to a
low temperature Debye temperature $\Theta{_D}(0)$ of $\approx$ 776(14)K
in good agreement with the results of the early heat capacity
study by Swift and White and recently by Bud'ko et
al.\cite{Swift,Budko}.

From a comparison with the results of various band structure
calculations, which consistently predict a Sommerfeld term $\gamma_{\rm
BS}\approx$1.7 mJ/mol\,K$^2$ \cite{Kortus,Kong,An}, an electron phonon
coupling parameter $\lambda_{\rm el-ph}$ is derived to

\begin{equation}
\lambda_{\rm el-ph} = \gamma_{\rm exp}/\gamma_{\rm BS} - 1 \approx
0.82(6)
\end{equation}

which is in very good agreement with theoretical predictions
\cite{Kong,Bohnen,Liu}.

Our results clearly substantiate the scenario of MgB$_2$ being in
the `intermediate or moderately strong' coupling regime as
conjectured from tunneling spectroscopy experiments by Sharoni et
al. \cite{Sharoni} and  more recently by Raman spectroscopy
\cite{Chen} and full-potential LMTO density-functional calculations
\cite{Kong}. The heat capacity measurements  do not suffer from
the inherent problems of tunneling or optical techniques, namely
surface defects or deterioration and our results indicate a gap
value of 7.0(3) meV for MgB$_2$. The ratio ($T_{\rm C}/\Theta_{\rm
D}$)$^2$ amounts to 2.3$\times$10$^{-3}$ and is close to classical
elemental strongly-coupled superconductors. From McMillans's
expression using  a typical value $\mu ^*\approx0.15$ for the
Coulomb pseudopotential we estimate an empirical electron-phonon
coupling constant $\lambda_{\rm el-ph}\,\approx\,1$ which is
somewhat larger than the value obtained from the linear heat
capacity term.

Since initial submission of this paper two more heat capacity
studies have been reported, both carried out on commercial samples
with 98\% nominal purity \cite{Ott,Junod}. Both papers find
somewhat decreased $T_{\rm C}$'s but within error limits identical
$\Delta c_P(T_{\rm C})$'s of $\approx$ 80 mJ/mol\,K$^2$. The work by
Wang et al. reliably derives $\gamma_{\rm exp}$=2.67 mJ/mol\,K$^2$
from measurements in magnetic fields up to 16 T which is  close to
our findings. From a subtraction of a theoretically estimated
lattice heat capacity from zero-field data W\"alti et al. find an
increased $\gamma_{\rm exp}$ of 5.5 mJ/mol\,K$^2$. As in our work
these investigations observe as well the striking difference
between $\gamma_{\exp}$ and $\gamma_{\rm fit}$ which now appears
to have become a robust result. Wang et al. assign this
discrepancy to possible non-BCS behavior resulting from gap
anisotropy and only partial condensation of the electrons at
$T_{\rm C}$. Liu et al. propose  multigap superconductivity and an
impurity-sensitive specific-heat jump at $T_{\rm C}$ as a possible
explanation \cite{Liu}.

In summary, we present and analyze heat capacity measurements
which show that the new  superconductor MgB$_2$ is a BCS
superconductor in the intermediate or 'moderately' strong
electron-phonon coupling regime. From fits of the heat capacity
jump we conclude 2$\Delta(0)/k_{\rm B}T_{\rm
C}$=4.2(2) corresponding to a gap of 7.0(3) meV at $T$=0\,K.

We would like to thank E.\ Br\"ucher and  G. Siegle for
experimental assistance. We gratefully acknowledge useful
discussions with O. K. Andersen, A. Bussmann-Holder, O. Dolgov, U.
Habermeier, O. Jepsen, J. K\"ohler, J. Kortus, and W. Schnelle.

\newpage

\begin{figure}
\centerline{\psfig{figure=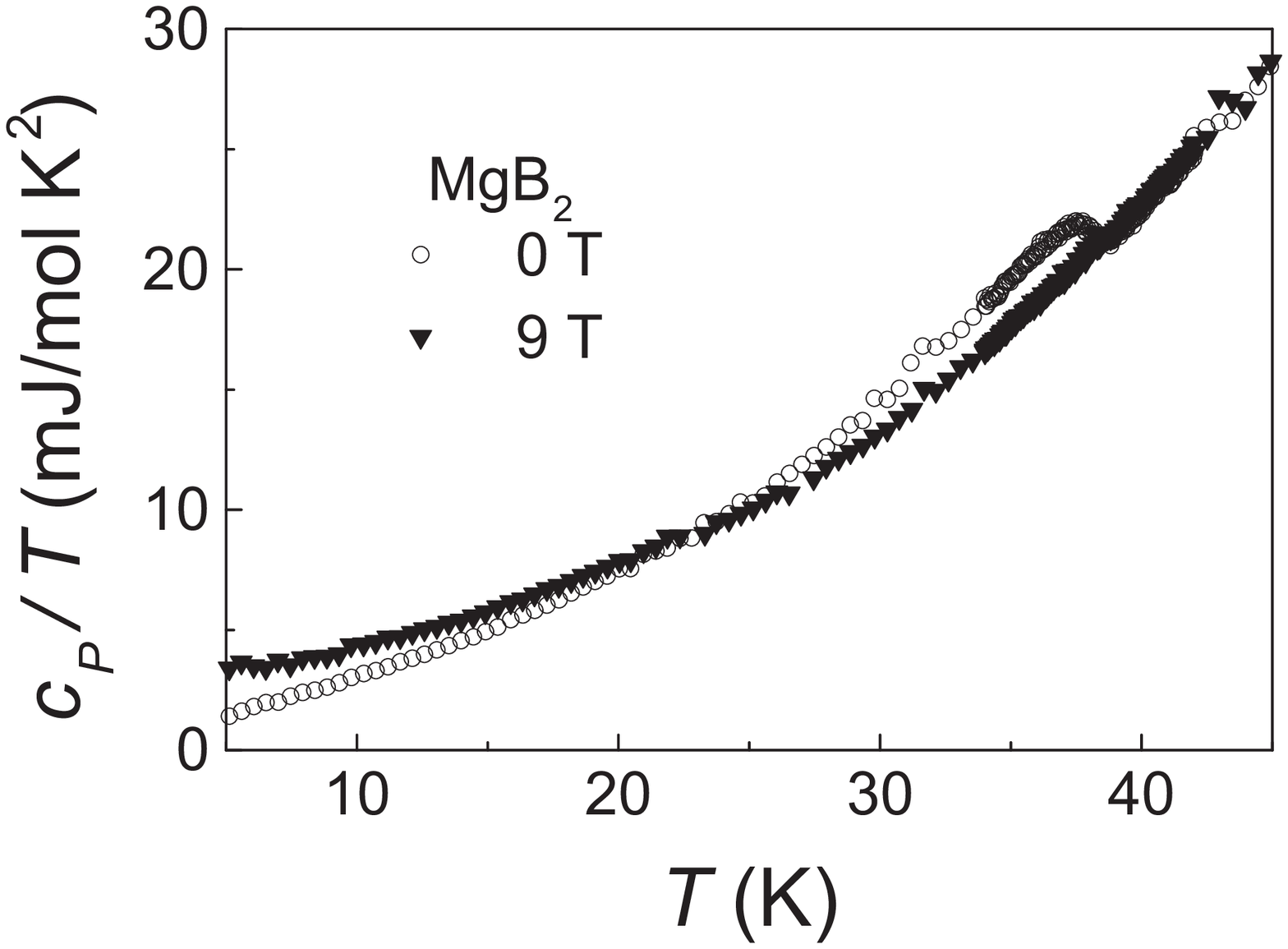,width=9cm,angle=0}} \caption{
Heat capacity $C_P/T$ of a polycrystalline sample of MgB$_2$
determined in zero field and with an applied magnetic field of 9 T
(sample 2).}
 \label{fig1}
\end{figure}

\begin{figure}
\centerline{\psfig{figure=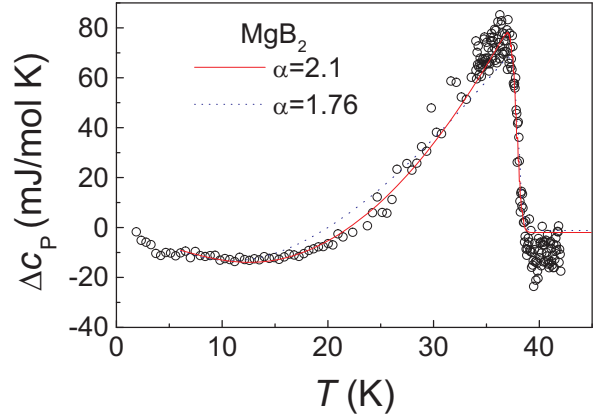,width=9cm,angle=0}} \caption{
Heat capacity  difference $\Delta c_P$ obtained as described in
the text. The full line represents the results of fits assuming
$\alpha$=$\Delta(0)/k_{\rm B}T_{\rm C}$=2.1 and a Sommerfeld
coefficient of $\gamma_{\rm fit}$ = 1.1 mJ/mol K$^2$. The dashed
curve indicates a fit with the weak coupling BCS heat capacity
($\alpha_{\rm BCS}$=1.76). Both fits allowed for a slight
temperature independent shift of the background ($<$ 2 mJ/mol\,K)
as described in the text.} \label{fig2}
\end{figure}

\begin{figure}
\centerline{\psfig{figure=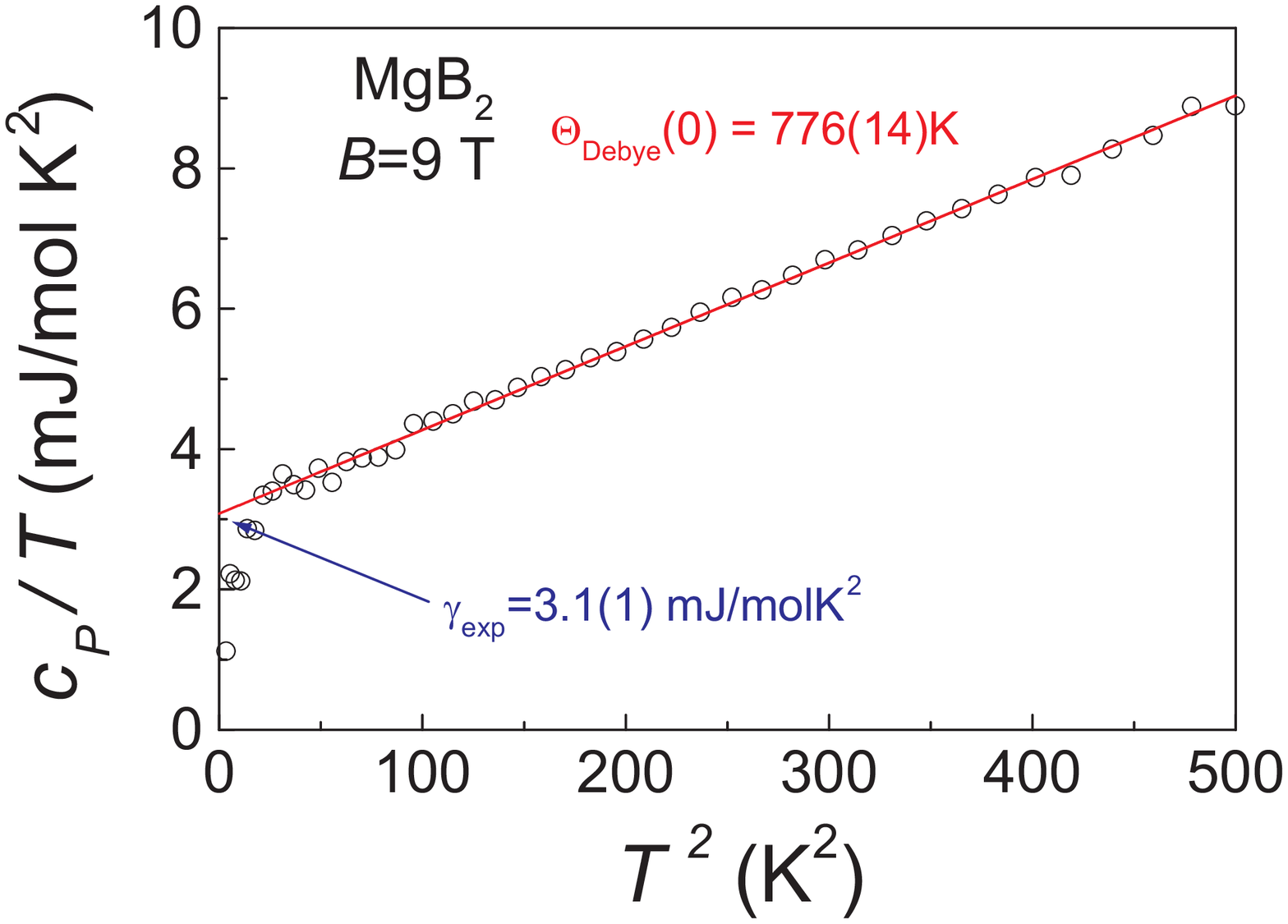,width=9cm,angle=0}} \caption{
Heat capacity $c_P/T$ versus $T^2$ together with a linear
approximation for temperatures $T >$5\,K (solid line). A
Sommerfeld coefficient $\gamma_{\rm exp}$ of 3.1(1) mJ/mol\,K$^2$
is obtained from the intersection with the ordinate.} \label{fig3}
\end{figure}


\begin{thebibliography}{99}
\bibitem{addenda}
If addenda taken at zero field are used as references for high
field measurements we observe a pronounced underestimation of the
linear term in the low-$T$ heat capacity. Consequently
$\gamma_{\rm exp}$ as extrapolated from $c_P/T$ versus $T^2$ plots
is found to be too low.
\end{thebibliography}
\end{document}